\documentclass[aps,prx,twocolumn,showpacs,groupedaddress,superscriptaddress,footinbib,longbibliography]{revtex4-2}

\usepackage{pdfpages} 
\usepackage{pgffor} 

\makeatletter
\AtBeginDocument{\let\LS@rot\@undefined}
\makeatother

\def\supplementfilename{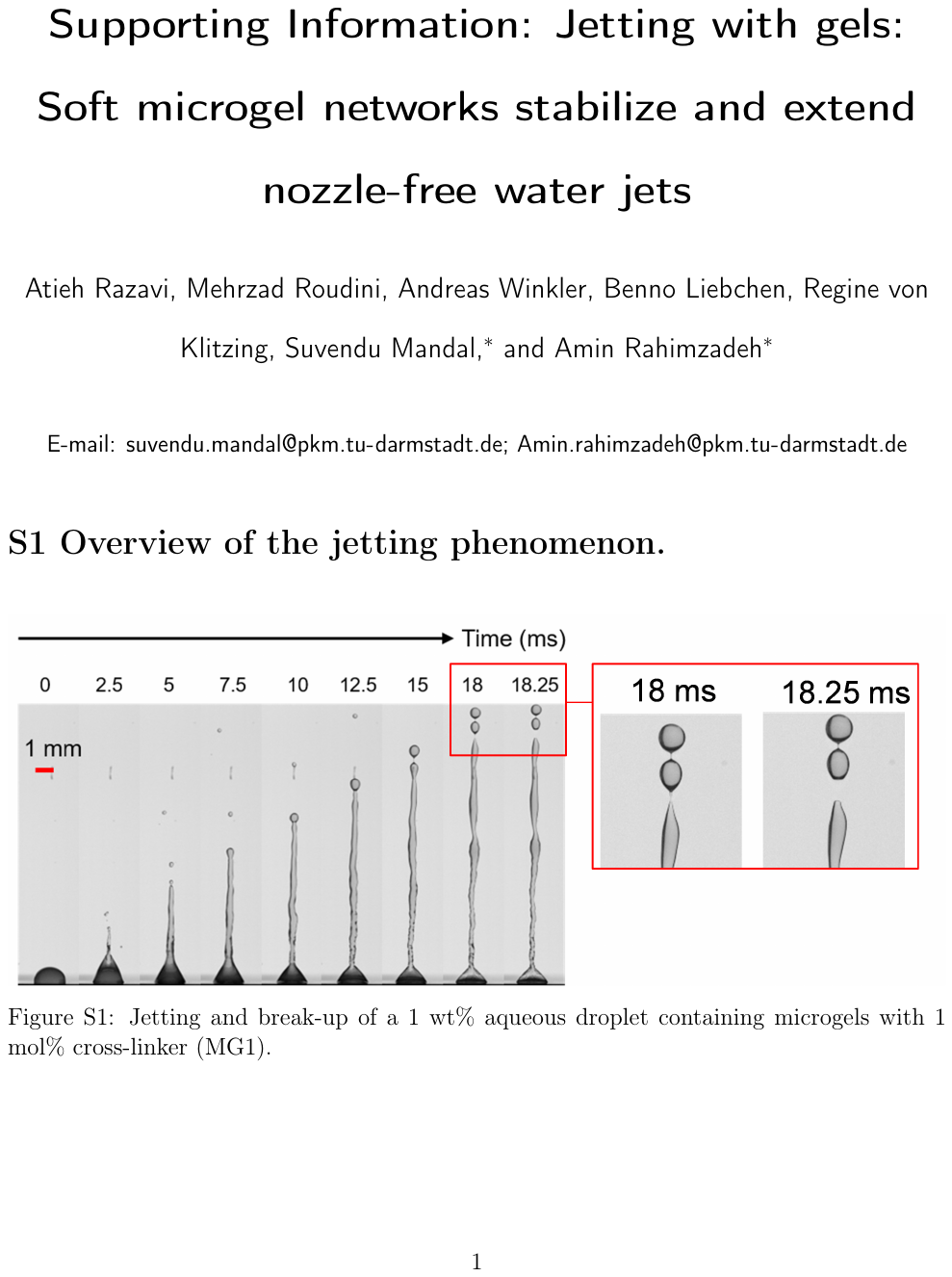}

\pdfximage{\supplementfilename}
\def\numbersupplementpages{\the\pdflastximagepages}

\newif\ifarXiv
\arXivtrue 
\usepackage{textcomp}

\usepackage{mathrsfs}
\usepackage{natbib,hyperref}
\setcitestyle{square,sort&compress,comma,numbers}
\hypersetup{colorlinks=true, citecolor=blue, urlcolor=blue, linkcolor=blue}
\usepackage[english]{babel}
\usepackage[utf8]{inputenc}
\usepackage{graphicx}
\usepackage[caption=false]{subfig}
\usepackage{amsmath}
\usepackage{amsfonts}
\usepackage{amssymb}
\usepackage{color,soul}
\setulcolor{red}
\usepackage{easyReview}
\usepackage{xcolor}

\usepackage[normalem]{ulem}

\usepackage{newunicodechar}

\begin{document}

\title{Jetting with gels: Soft microgel networks stabilize and extend nozzle-free water jets}
\author{Atieh Razavi}
\affiliation{Soft Matter at Interfaces, Department of Physics, Technische Universität Darmstadt, Hochschulstraße 8, 64289 Darmstadt, Germany}
\author{Mehrzad Roudini}
\affiliation{IFW Dresden, SAWLab Saxony, Acoustic Microsystems, Helmholtzstr. 20, 01069 Dresden, Germany}
\author{Andreas Winkler}
\affiliation{IFW Dresden, SAWLab Saxony, Acoustic Microsystems, Helmholtzstr. 20, 01069 Dresden, Germany}
\author{Benno Liebchen}
\affiliation{Technische Universit\"at Darmstadt, Karolinenplatz 5, 64289 Darmstadt, Germany}
\author{Regine von Klitzing}
\affiliation{Soft Matter at Interfaces, Department of Physics, Technische Universität Darmstadt, Hochschulstraße 8, 64289 Darmstadt, Germany}
\author{Suvendu Mandal}
\email{suvendu.mandal@pkm.tu-darmstadt.de}
\affiliation{Technische Universit\"at Darmstadt, Karolinenplatz 5, 64289 Darmstadt, Germany}
\author{Amin Rahimzadeh}
\email{amin.rahimzadeh@pkm.tu-darmstadt.de}
\affiliation{Soft Matter at Interfaces, Department of Physics, Technische Universität Darmstadt, Hochschulstraße 8, 64289 Darmstadt, Germany}

\begin{abstract}
The stability of high-speed liquid jets is crucial for applications ranging from precision printing to needle-free drug delivery, yet it is fundamentally limited by capillary-driven breakup. A common strategy to stabilize jets is to use surfactants to lower surface tension. However, in nozzle-free jetting driven by surface acoustic waves (SAWs), extreme deformation rates cause conventional surfactants to desorb, calling for alternative strategies to stabilize jets. In particular, we find that tuning the nanoscale softness of PNIPAM microgels provides a robust, biocompatible strategy to overcome this limitation. Soft, low–cross-linker-density microgels form elastic interfacial networks at the air–water interface that suppress surface tension recovery, delay Rayleigh–Plateau instabilities, and extend SAW-driven jet lengths by up to $44\%$. In contrast, stiffer microgels lose network cohesion under strain, leading to rapid jet breakup. To gain molecular-level insights, we perform dissipative particle dynamics simulations, which reveal that polymer bridges in soft microgels remain entangled during elongation, maintaining a reduced effective surface tension. Finally, a simple scaling analysis, balancing the SAW-driven kinetic energy imparted to the droplet against the surface energy required to form a jet, quantitatively predicts the observed length enhancement. This surfactant-free, biocompatible approach lays the foundation for long-lived jets, enabling precision needle-free drug delivery, high-speed printing, and other high-strain interfacial flow applications.
\end{abstract}

\maketitle


\section{\label{sec:level1}Introduction}
Fluid jets~\cite{Jordan:2022PRF, Deblais:PRL2018, Utada:2007PRL,Utada:PRL2008, Dong2024,Challita:2024AnnRev,Challita:2024PNAS,Kushwaha:PRL2025} are  traditionally generated by forcing liquid through a nozzle—commonly seen in syringes, inkjet printers~\cite{Hoath2012, Antonopoulou2021, Lohse2022}, or spray systems~\cite{Liu:NatCommn2023}. Here, we exploit a nozzle-free jetting approach based on highly focused surface acoustic waves (SAWs), which can be viewed as nanoscale analogues of seismic waves traveling along the Earth’s surface. This method destabilizes the liquid-air interface, overcoming capillary stresses to generate a collimated liquid stream known as a liquid jet \cite{Eggers_2008}. These jets serve as a paradigm for free-surface motion, capable of carrying materials within the liquid and transferring them to the intended site. However, a fundamental limitation of SAW-induced jetting is its susceptibility to capillary-driven Rayleigh–Plateau instabilities, which lead to premature jet breakup and constrain jet length \cite{Eggers_2008}. These instabilities pose a major challenge in high-speed applications, where jet stability is essential for accurate droplet deposition, transdermal drug penetration, and microfabrication processes. Despite its technological promise, most research on SAW-driven jetting has focused on pure liquids \cite{Tan2009, Lei2020}, the role of interfacial stabilizers in this context remains poorly understood.

Previous studies in air-blast jetting have shown that surfactants can increase jet breakup length by lowering surface tension and inducing Marangoni stresses that counteract capillary-driven instabilities \cite{Wang2025, Antonopoulou2021, Shavit1995, Christanti2001, Li2023, Sijs2021, Dechelette2011}. In free-falling water jets into toluene, surfactant transport across the interface delays jet breakup and increases jet length \cite{Toor:2017NanoLetters}. Yet in SAW-driven systems, interfacial deformation occurs so rapidly that surfactants desorb before providing lasting stabilization. Biocompatibility concerns further limit their use in biomedical contexts, motivating the search for an alternative strategy capable of withstanding high strain rates while remaining non-toxic.

A promising alternative is soft colloidal stabilizers, particularly microgels—cross-linked polymer networks swollen by solvent. Unlike rigid colloidal particles, microgels are highly deformable, dynamically adapting at liquid interfaces through adsorption, spreading, and steric interactions \cite{Destribats11, minato18, Rey20, Scheffold2020, FERNANDEZ2021, stockmdpi22, kuhnhammer22,Grillo:2020Nature,Grillo:2020Nature,Volk:2019PCCP,Menath:2021PNAS,Scotti:2019NatCommn,Schmidt:2023PRL,Brito:2024JCP,DelMonte:2024PRX}. Their distinctive “fried-egg” morphology features a swollen corona that spreads across the interface while a denser core remains mainly submerged, with a small portion protruding into the air phase \cite{Geisel:2012unraveling,Camerin:2020PRX, Hazra2024}. Although microgels are well established for stabilizing emulsions and coatings~\cite{Rey:2023NatureCommn, Chen:2016}, their behavior under fast, high-strain interfacial flows such as SAW-driven jetting has not been investigated.

Here, we show that the softness of poly(\emph{N}-isopropylacrylamide) (PNIPAM) microgels plays a decisive role in stabilizing nozzle-free jets. Microgels with low cross-linker density assemble into entangled, elastic interfacial networks that suppress Rayleigh–Plateau instabilities, producing jets that persist up to $44\%$ longer than those from pure water. In contrast, stiffer microgels lose network connectivity under strain, allowing the air–water surface tension ($\approx 70$ mN m$^{-1}$) to recover and prompting earlier jet breakup. Due to experimental challenges in capturing dynamic surface tension during jetting, we employ dissipative particle dynamics (DPD) simulations with explicit solvent to gain molecular-level insights. DPD simulations reveal that polymer chains bridging neighboring soft microgels remain entangled during high-speed deformation, preserving a lower surface tension than the pure air–water interface throughout jet elongation. Subsequently, we develop a simple scaling argument, balancing SAW-driven kinetic energy with the jet’s surface energy, which quantitatively predicts the observed length enhancement.

By combining experiments, simulations, and scaling analysis, we establish a direct link between nanometer-scale polymer architecture and centimeter-scale jet stability. This surfactant-free, biocompatible strategy enables long-lived jets, opening avenues for precision liquid delivery, high-speed printing, and dynamic control of high-strain interfacial flows.

\section{Results}

\subsection{Controlled jetting using surface acoustic waves}
\begin{figure*}
\includegraphics[width=1\textwidth]{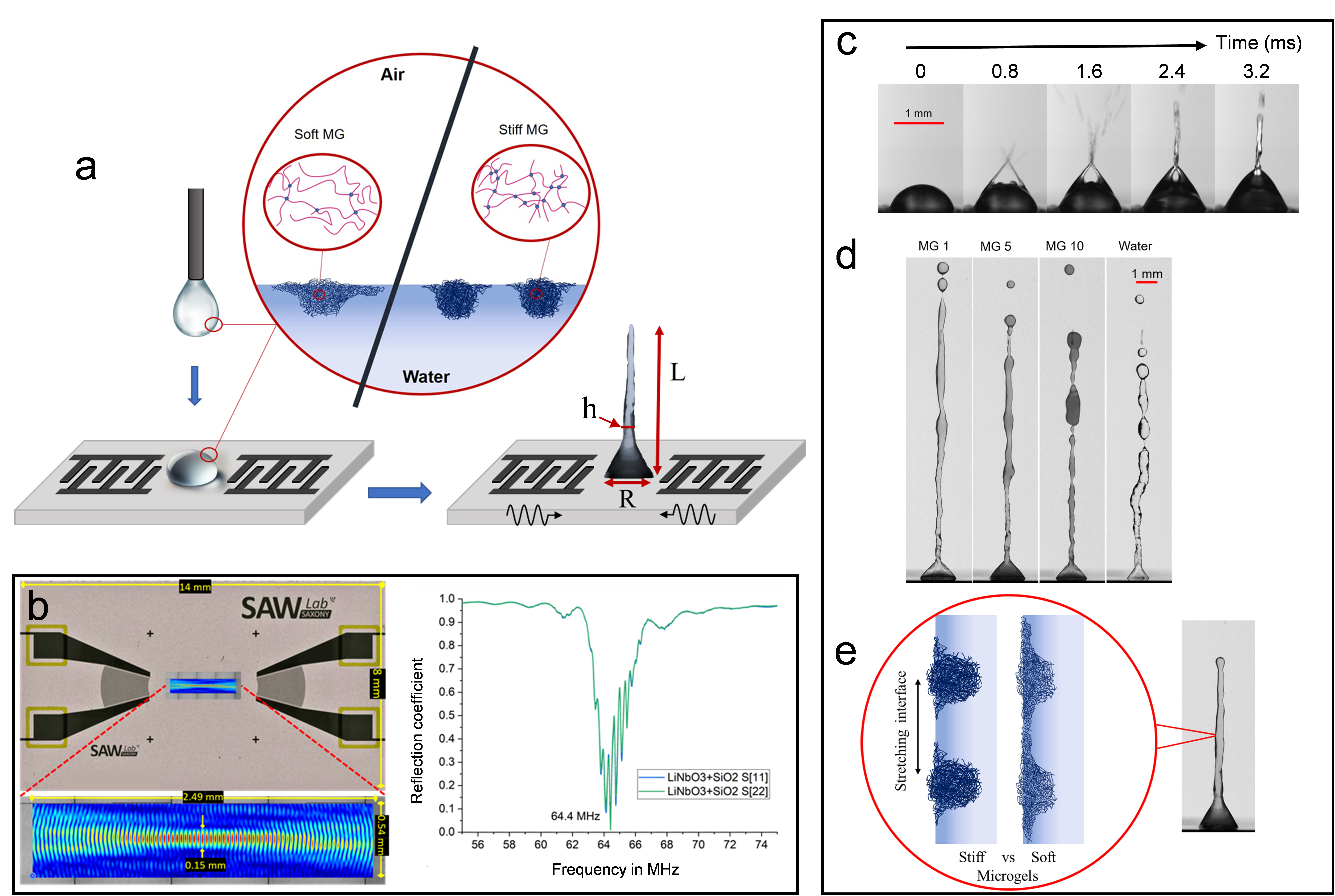}
\caption{\label{fig:combi} \textbf{SAW-driven jetting of microgel dispersions.} 
(a) An aqueous droplet containing microgels of varying concentration and stiffness (cross-linker density) is gently placed at the centre of a SAW chip between two opposing interdigital transducers (IDTs). After allowing microgels to adsorb and equilibrate at the interface, a focused SAW field drives jet formation with diameter $h$ and length $L$, recorded via high-speed imaging; jet performance is quantified as $L/R$, where $R$ is the initial droplet radius. (b) Top view of the interdigital transducer (IDT) and laser Doppler vibrometry (LDV) measurement showing the acoustic wave field profile superimposed on microscope images of the SAW chip. The standing surface acoustic wave (SAW) is generated at a wavelength of 60$~\mu$m and an excitation frequency of 64.4 MHz according to electrical characterization of the standing surface acoustic wave chip in terms of its reflection coefficient magnitude (\(|S11|\)). (c) In the early stages of jet formation, surface destabilization of the droplet (here water) occurs at two symmetric locations due to oppositely traveling surface acoustic waves (SAWs) of equal energy, leading to the formation of a vertical jet. A similar process is observed in microgel dispersions. (d) The maximum jet length of water and 1 wt\% microgel dispersions at different cross-linker (BIS) contents (MG1: 1 mol\%, MG5: 5 mol\% and MG10: 10 mol\%). (e) Schematic representation of a possible scenario of how soft vs stiff microgels behave at the interface during its extension. }
\end{figure*}

The SAW device comprises slowness-curve–adjusted interdigital transducers (IDTs)\cite{o2020slowness} arranged in a delay-line configuration to generate a strong standing wave at the chip center. This geometry focuses the acoustic energy into a well-defined interaction zone, only $1.6\lambda_{\mathrm{SAW}}$ (FWHM) wide and approximately $25\lambda_{\mathrm{SAW}}$ (FWHM) long, ensuring highly localized coupling between the SAW field and the fluid above. A representative microscopic image of the chip layout, overlaid with laser Doppler vibrometry (LDV) measurements of the amplitude distribution at the optimal driving frequency [see Fig.~\ref{fig:combi}b].

We investigate the effect of microgel softness on SAW-driven jet formation by preparing droplets of varying volumes from microgel dispersions of three stiffness levels with 1\%, 5\% and 10\% of nominal cross-linker concentrations (MG1, MG5, and MG10, with MG1 being the softest) at dispersion concentrations of $0.01$, $0.1$, and $1$ wt$\%$. These droplets are dispensed using a pipette [see Fig.~\ref{fig:combi}a] and placed on a SAW chip. The dispensed droplet is equilibrated for 2–3 minutes prior to IDT activation. It is known that microgels adsorb at the air-water interface~\cite{Tatry2023} (prior to jetting), and that lower cross-linker density creates softer microgels, while higher cross-linker density produces stiffer ones~\cite{Burmistrova2011} [see Fig.~\ref{fig:combi}a]. Upon activating the transducer at a combined power of 4 W, SAWs induce jetting, and the jet length ($L$) relative to the initial droplet baseline ($R$), as shown in Fig.~\ref{fig:combi}a, is recorded using a high-speed camera.

During the early stages of jet formation which takes less than 3 ms [see Fig.~\ref{fig:combi}c], the counter-propagating SAWs transfer energy into the sessile droplet, destabilizing the surface at two symmetric locations, in agreement with theoretical predictions~\cite{Lei2020}. This instability arises from two equal streaming forces generated by high acoustic pressure, known as \textit{caustics}~\cite{Riaud2017}, whose superposition leads to the formation of a single jet ejected perpendicularly to the substrate. During the jet formation, satellite droplets may form and disintegrate, but they are excluded from the analysis. The jet initially ejects when the inertial forces overcome capillary stresses at the interface, launching with a certain velocity~\cite{Tan2009}. It subsequently undergoes a classical Rayleigh-Plateau instability in which surface tension drives breakup of the jet into multiple droplets.

\subsection{Softness-induced jet stability}

Jet stability is governed by the balance of inertial, viscous, surface tension, and gravitational forces. By fixing the input IDT power at $4$ W for all samples (jet velocity of about $1$ m/s), we hold inertial forces constant, enabling direct comparison of remaining forces (viscous, surface tension, and gravitational). At this velocity, we estimate the gravitational length—the characteristic scale at which gravity balances inertial acceleration—by equating convective inertia and gravitational acceleration: $|\mathbf{v} \cdot \nabla \mathbf{v}| \sim u_0^2/\ell_g \sim g$. This yields a gravitational length $\ell_g = u_0^2/g \approx 10$ cm, which far exceeds the maximum jet lengths observed in our experiments. Furthermore, we estimate the capillary length, $\ell_c = \sqrt{\gamma / (\rho g)}$, which defines the length scale at which surface tension balances gravity. For our system, $\ell_c$ ranges from approximately $2.1$ mm in microgel dispersions [see Fig. S2] to $2.7$ mm in pure water, values that are significantly larger than the jet radius ($\approx 100$ $\mu$m). Viscous effects are also comparable across the samples [see Fig. S3]. Therefore, we conclude that capillary forces, rather than gravitational or viscous forces, dominate the interfacial dynamics that govern jet stability and breakup in our system.

\begin{figure}
\includegraphics[width=0.5\textwidth]{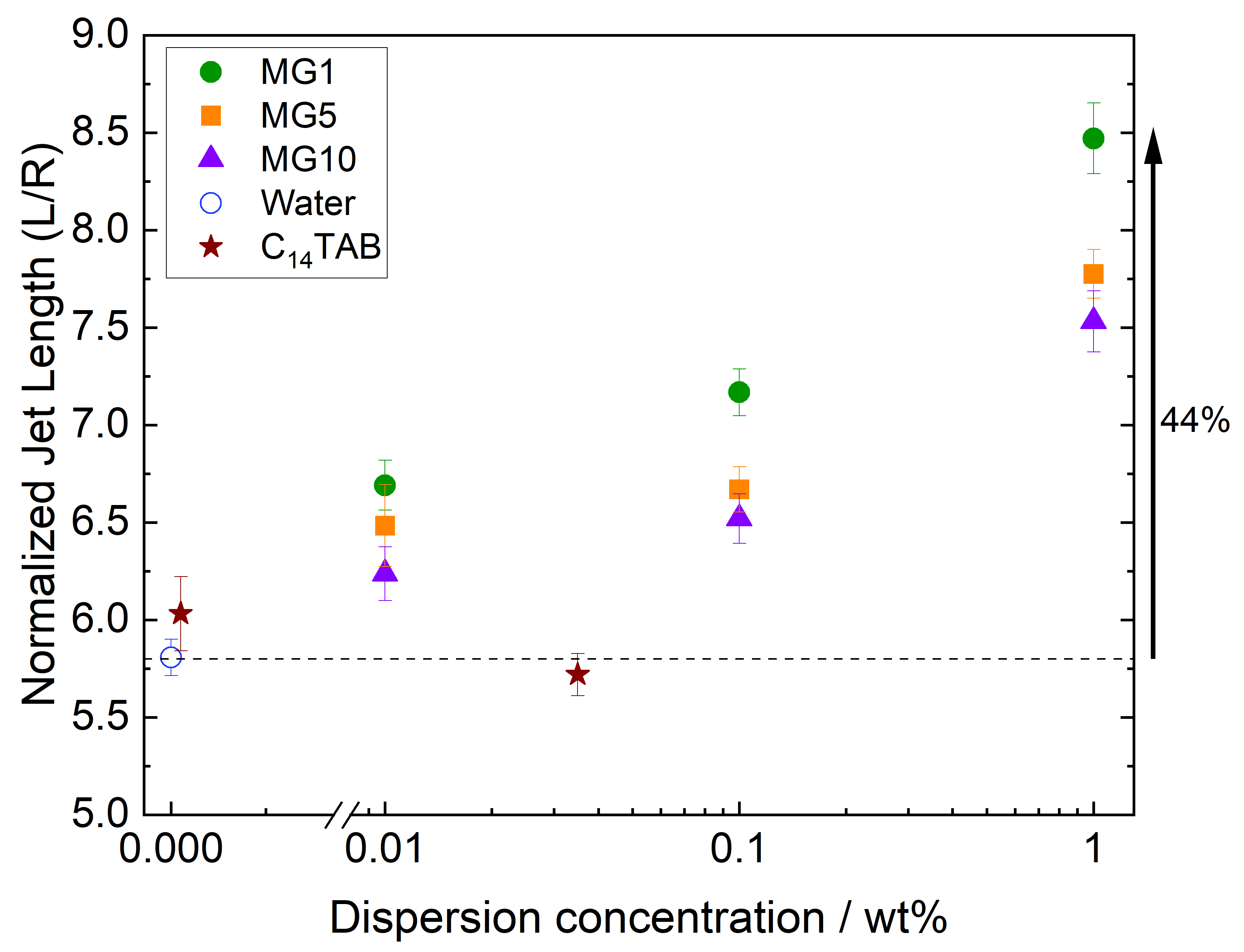}
\caption{\label{fig:jet-L} \textbf{Effect of concentration and interfacial stabilizer type on normalized jet length.} Maximum jet length normalized by the droplet's initial baseline ($L/R$) as a function of liquid concentration. Data are shown for pure water (concentration = 0, also showed with the dasshed line), C$_{14}$TAB solutions at two different concentrations of about $0.01$ and $0.3$ CMC, and microgel dispersions with varying stiffness (different cross-linkerdensities): MG1, MG5, and MG10, across their respective aqueous concentrations.}
\end{figure}

We observe that introducing PNIPAM microgels into aqueous droplets enhances jet stability, as evidenced by increased jet length compared to pure water [see Fig.~\ref{fig:combi}d]. For example, the jet from a 10 $\mu$l droplet containing soft microgels (MG1) at 1 wt\% concentration reaches a maximum length of $15.59 \pm 0.14$ mm at 18 ms and breaks up at 18.25 ms [see Fig.~S1]. A pronounced dependence of jet behavior on microgel softness is observed, with soft microgels (e.g., MG1) markedly suppressing droplet breakup relative to their stiffer counterparts (e.g., MG10), leading to longer jets. We hypothesize that the extensive deformability of soft microgels at the jet surface facilitates the formation of a cohesive interfacial layer, delaying exposure to the bare air–water interface [see Fig.~\ref{fig:combi}e, Movies Water.avi and Microgel.avi].
To quantify this effect, we measure the maximum jet length $L$, normalized by the droplet base-line diameter $R$ (denoted $L/R$), across a range of dispersion concentrations. Even at low concentrations (0.01~wt\%), microgels increase $L/R$ relative to pure water [Fig.~\ref{fig:jet-L}], with the enhancement becoming more pronounced as both softness and concentration increase. At 1~wt\%, soft microgels (MG1) extend the jet length by approximately 44\% compared to water. At lower concentrations, the stabilizing effect diminishes, likely due to insufficient interfacial coverage prior to breakup. This concentration-dependent enhancement suggests that sufficient interfacial coverage by soft microgels is critical to achieving maximal jet stability.

\begin{figure*}
\includegraphics[width=1\textwidth]{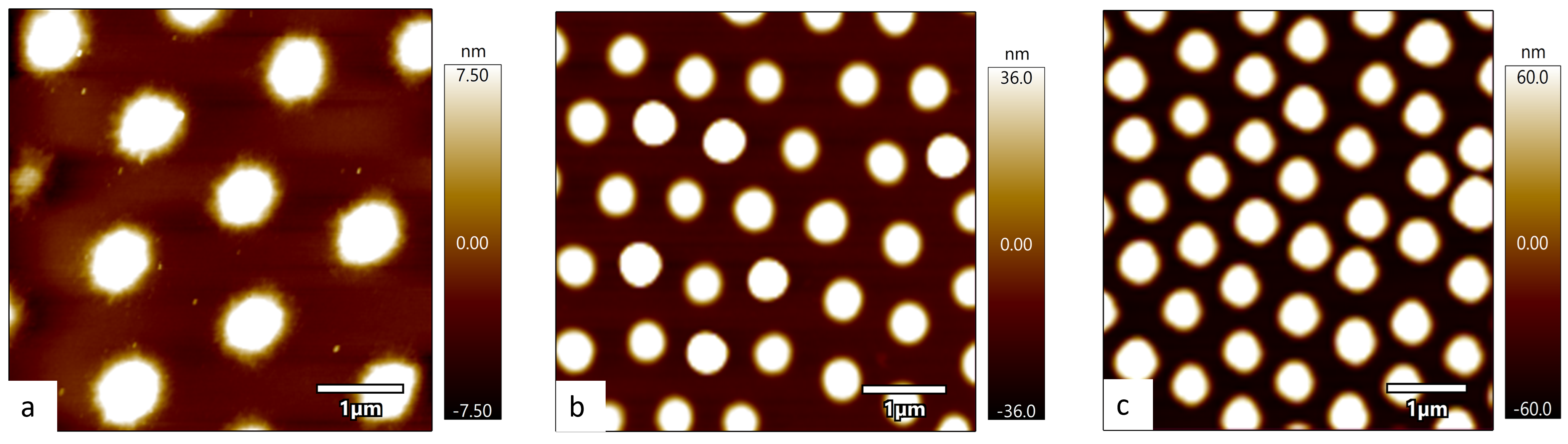}
\caption{\label{fig:langmuir} \textbf{\textit{Ex-situ} AFM characterization of microgel arrangements at the air–water interface.} The structure of the microgel particles at constant lateral pressure through the AFM at $5 \, \mu \text{m}^2$ scan. The layers were transferred to a silicon wafer using a Langmuir trough method at a lateral pressure of 5mN/m for (a) MG1 (b) MG5 (c) MG10, The scales differ because the height of each scan varies significantly.}
\end{figure*}

To understand why softer microgels generate longer jets and delay breakup, we investigate their interfacial organization using Langmuir–Blodgett deposition (see methods) under a fixed lateral pressure of 5~mN/m, selected to approximate conditions during jetting. As shown in Fig.~\ref{fig:langmuir}, softer microgels form flatter, more extended monolayers that efficiently cover larger interfacial areas with fewer particles than their stiffer counterparts. Although the dangling polymer chains of the microgels are not resolved in our AFM measurements, it is known that microgels interact through their outer, loosely cross-linked coronas \cite{Tatry2023}, which likely remain in contact at the interface. This efficient packing facilitates faster interfacial coverage and allows surface tension to reach a reduced value more rapidly compared to stiffer microgels \cite{StockADVSC}, as confirmed by pendant drop measurements [see Fig.~S2]. While direct measurement of surface tension during jet elongation remains experimentally challenging, our molecular dynamics simulations support this scenario, showing that soft microgels remain entangled and connected under rapid interfacial extension (see Modeling section).

\begin{figure}
\includegraphics[width=0.5\textwidth]{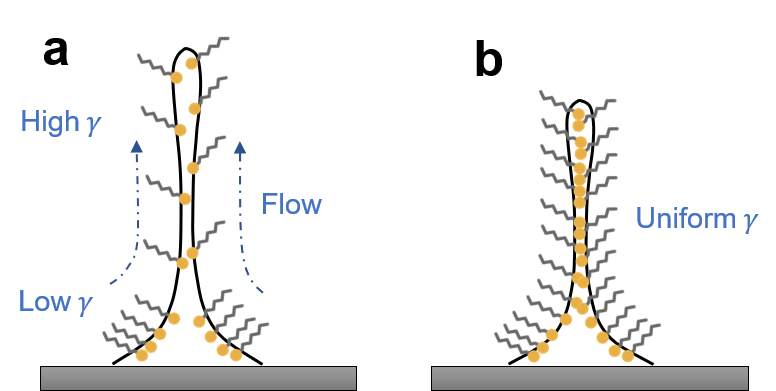}
\caption{\label{fig:surfactant}  \textbf{Schematic representation of a jet from a sub-CMC surfactant solution.} (a) At low surfactant concentrations (sub-CMC), interfacial extension generates gradients in surfactant coverage, driving Marangoni flows that transiently counteract breakup and extend jet length.
(b) At high concentrations (near CMC), the interface is uniformly saturated, eliminating surface tension gradients and suppressing Marangoni flows, which leads to reduced jet stability and shorter jets.}
\end{figure}

To demonstrate that conventional surfactants are less effective than microgels under identical jetting conditions (same speed $u_0$ and droplet size $R$), we tested C$_{14}$TAB at both sub-critical ($0.01$ CMC) and near-CMC ($0.3$ CMC) concentrations. We find that at low concentrations, C$_{14}$TAB provides a modest improvement ($\sim 4\%$) in jet length compared to pure water. We anticipate that it occurs due to Marangoni flows that transiently resist breakup [see Fig.~\ref{fig:surfactant}a]. However, as concentration increases toward CMC, jet length decreases slightly—suggesting that excess surfactant suppresses Marangoni flows, leading to faster destabilization and earlier breakup [see Fig.~\ref{fig:surfactant}b]. This behavior is also consistent with recent findings in air-blast jetting \cite{Wang2025}. In contrast, microgels—particularly soft ones—form cohesive, stretchable \textit{interfacial networks} that sustain reduced surface tension during rapid deformation, a mechanism not achievable with molecular surfactants.

\begin{figure*}
\includegraphics[width=1\textwidth]{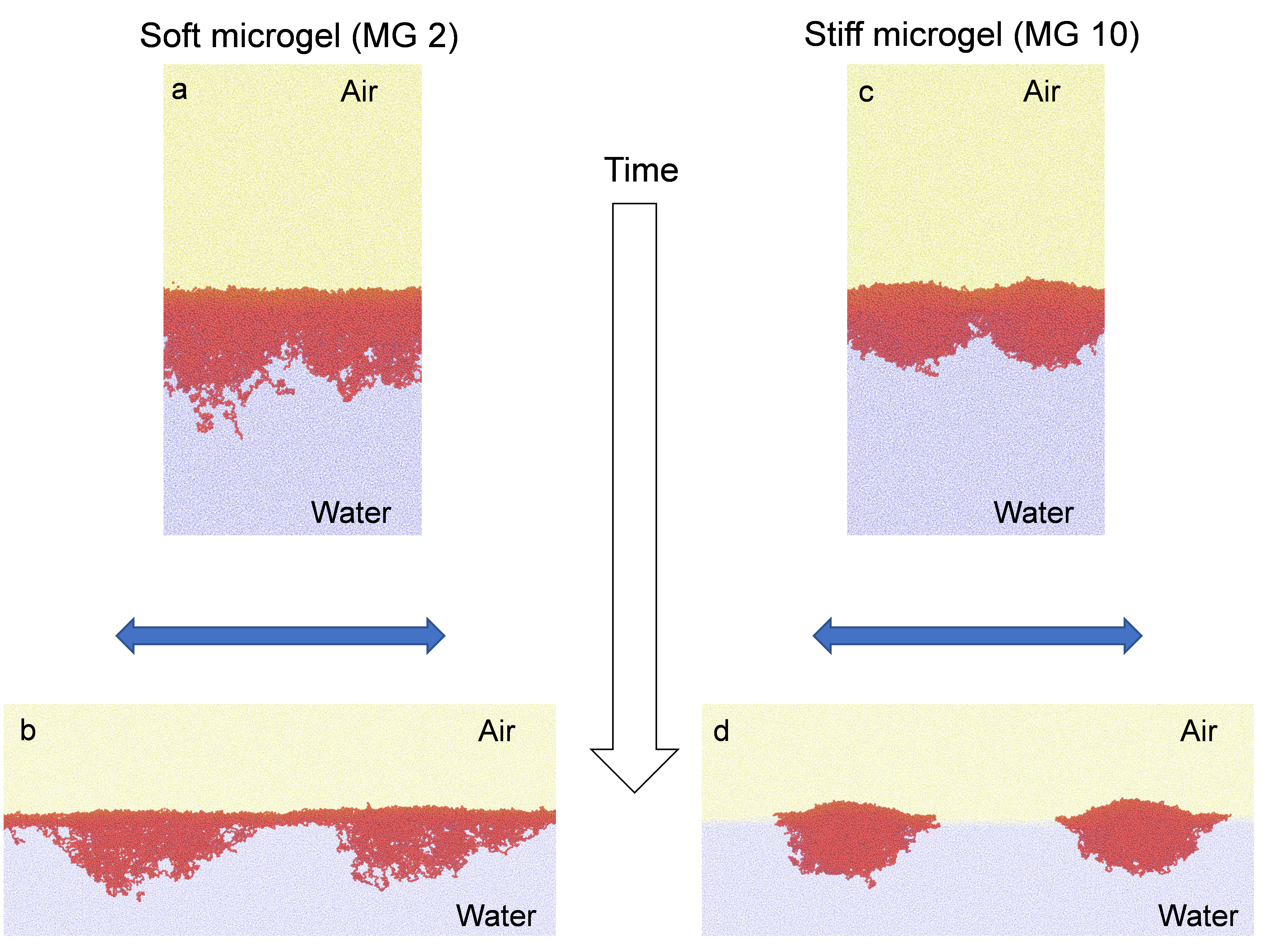}
\caption{\label{fig:snap_ini} \textbf{Simulation snapshots of soft and stiff microgels at the air-water interface.} Representative configurations of soft (a, b) and stiff (c, d) microgels (red beads) adsorbed at the air–water interface. Initially, two microgels are positioned in close lateral contact (a, c). The simulation box is then stretched laterally at constant volume to mimic interfacial extension during jetting, resulting in pronounced deformation and persistent connectivity for soft microgels (b), in contrast to the rapid disentanglement observed for stiff microgels (d).}
\end{figure*}

\begin{figure}
\includegraphics[width=0.5\textwidth]{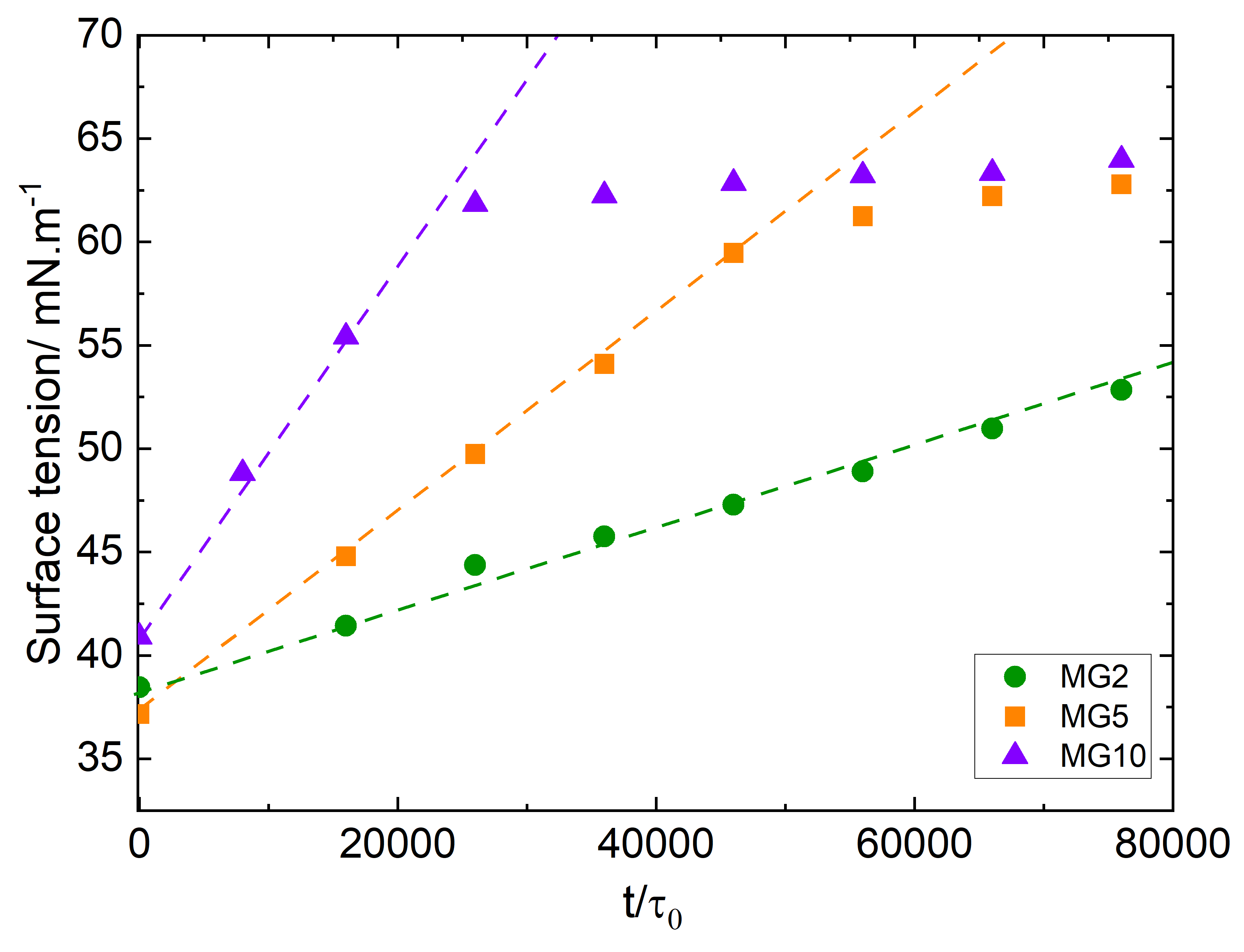}
\caption{\label{fig:surface_tension} \textbf{Simulated surface tension dynamics.} Dynamic surface tension as a function of time for microgels with different cross-linker densities. The plot illustrates how varying the cross-linker concentration influences the rate at which the surface tension evolves, highlighting the distinct behaviors of soft (low cross-linker density) and stiff (high cross-linker density) microgels during interfacial stretching.}
\end{figure}

\subsection{Modeling the stretching of the air-water interface}
To complement our experimental investigation of jet stabilization by microgels under SAW actuation, we developed a coarse-grained simulation framework to probe the molecular-scale mechanisms governing microgel behavior at dynamically deformed air–water interfaces. Our model mimics the stretching of a fluid jet by placing two microgels at an initially equilibrated interface and imposing lateral elongation at a constant velocity of $1$ m/s (see Methods), consistent with experimental jet elongation rates captured via high-speed imaging.

Microgels are constructed from a self-assembled network of monomer and cross-linker beads, enabling systematic variation in cross-linker density. We focus on two representative cases: soft microgels (MG2, $2\%$ cross-linker) and stiff microgels (MG10, $10\%$ cross-linker).  Due to slow equilibration of the $1\%$ cross-linked network system, MG2 was chosen as the softest microgel to complement experiments. The surrounding medium consists of a binary DPD fluid, representing water (blue) and air (yellow) phases. Following established approaches \cite{Gerelli:2024softness}, the microgels are first equilibrated at the interface. As expected, MG2 microgels exhibit extended, diffuse coronas with numerous dangling polymer chains, whereas MG10 microgels form compact, spherical structures with minimal surface flexibility [Fig.~\ref{fig:snap_ini}a and c].

Under imposed interfacial stretching, the two microgel types exhibit strikingly different responses. MG10 microgels quickly disentangle from one another, showing minimal resistance to separation. In contrast, MG2 microgels remain interconnected throughout the deformation process, facilitated by extensive chain interpenetration and high deformability [Fig.~\ref{fig:snap_ini}b and d]. This persistent adhesion allows soft microgels to maintain a contiguous interfacial layer even under dynamic strain [see Movie Soft.mp4].

This mechanical distinction has direct implications for interfacial properties. Prior to stretching, all systems—MG2, MG5, and MG10—exhibit comparable surface tension values of approximately 40 $\mathrm{mN\,m^{-1}}$, consistent with experimental observations [see Fig.~\ref{fig:surface_tension}; see also Fig. S2]. Upon initiating lateral deformation, the surface tension increases linearly with time as the microgels respond to interfacial strain, ultimately reaching a steady-state plateau. Notably, the temporal evolution of surface tension reveals stark contrasts between soft and stiff microgels: while MG10 systems rapidly recover toward the bare air–water tension of $\sim65\,\mathrm{mN\,m^{-1}}$ within $t/\tau_0 \approx 20000$, the softest microgels (MG2) maintain a substantially reduced surface tension, remaining below $50\,\mathrm{mN\,m^{-1}}$ up to $t/\tau_0 \approx 80000$ [Fig.~\ref{fig:surface_tension}]. These findings indicate that soft microgels can dynamically stabilize the interface under strong deformation.

\subsection{Scaling analysis: Conversion of kinetic energy to surface Energy}
To rationalize our experimental observations, we analyze the conversion of a millimeter-size droplet of initial diameter $R$ into a high-speed, collimated jet of diameter $h$ and length $L$ under surface acoustic wave (SAW) excitation [see Fig.~\ref{fig:combi}a]. In this nozzle-free and reservoir-free system, jet formation does not rely on continuous fluid supply; rather, it emerges solely from acoustic energy injected locally by the SAW. This energy is concentrated beneath the droplet, generating inertial streaming flows that drive fluid momentum toward the droplet’s free surface. When the flow inertia overcomes the stabilizing influence of surface tension, the liquid interface deforms and extrudes into a slender jet extending several centimeters—achieved without conventional nozzles or orifices.

Dimensional analysis~\cite{Tan2009,Challita:2024AnnRev} suggests that viscous and gravitational effects are negligible in this regime: the Capillary number, $\mathrm{Ca} = \mu u_0 / \gamma \sim 0.01$, represents the ratio of viscous stress ($\mu u_0 / h$) to capillary stress ($\gamma / h$) and indicates that viscous dissipation is weak; similarly, the Bond number, $\mathrm{Bo} = \rho g h^2 / \gamma \sim 0.001$, compares gravitational stress ($\rho g h$) to capillary stress ($\gamma / h$) and indicates that gravity has minimal influence. Under these conditions, jet dynamics are governed primarily by a balance between inertia and surface tension. Hence, the entire kinetic energy imparted by the SAW to the droplet, $E_{\mathrm{kin}} \sim  \rho \pi R^3 u_0^2/6$, is converted into the surface energy needed to form a cylindrical jet of diameter $h$ and length $L$, given by $E_{\mathrm{surf}} \sim \gamma \pi h L$. Equating these two energy scales ($E_{\mathrm{kin}} \sim E_{\mathrm{surf}}$) yields the scaling relation
\begin{equation}\label{eq:scalingEB}
\frac{L}{R} \sim \frac{\rho u_0^2 R^2}{6 \gamma h},
\end{equation}
which predicts that, for fixed $\rho$, $u_0$, $h$, and $R$, a reduction of the surface tension $\gamma$ produces longer jets. 

In our DPD simulations, soft microgel dispersions (e.g., MG2) lower the effective surface tension from $\sim 70$ mN m$^{-1}$ (pure water) to $\sim 50$ mN m$^{-1}$. This leads to a predicted jet elongation of about $40\%$, in excellent agreement with our experimental observations. Most importantly, using our experimental parameters ($\rho \approx 1000$ kg/m$^3$, $u_0 \approx 1$ m/s, $\gamma \approx 70$ mN/m, $R \approx 1$ mm, and $h \approx 0.2$ mm), the scaling relation predicts jet lengths on the order of $\mathcal{O}(1~\mathrm{cm})$, in agreement with our experimental data. This is an order of magnitude greater than the breakup length from the Rayleigh--Plateau instability~\cite{Eggers_2008}, $L_\mathrm{RP} \sim u_0 \left(\rho h^3/8\gamma \right)^{1/2} \sim \mathcal{O}(1~\mathrm{mm})$, demonstrating that our jets are not instability-limited. Instead, their length is set by the energy-balance scaling in Eq.~\ref{eq:scalingEB} as the governing mechanism for SAW-driven jetting.

\section{Discussion}
Our findings show that microgel softness, tuned via cross-linker density, governs interfacial resilience during rapid deformation. Unlike pure water jets, soft microgels form deformable interfacial networks that suppress Rayleigh–Plateau instabilities, extending jet lengths by up to $44\%$.

While our system centers on acoustically driven jets, the underlying mechanism likely extends to a wide range of dynamic interfacial processes. Similar mechanically stabilizing behaviors may occur in natural systems, such as bubble bursting at oil-covered water surfaces, where interfacial layers modulate jet formation \cite{Ji:2021compound}. In such scenarios, surface-active species—including lipids, proteins, and microorganisms—can form elastic barriers that mirror the stabilizing behavior of soft microgels. This analogy suggests that our findings may inform a broader understanding of how deformable interfacial structures influence jetting and breakup dynamics in both environmental and technological contexts.

Beyond their fundamental relevance, our results have direct implications for bioprinting and tissue engineering, where controlled jetting of soft, cell-compatible materials is essential. SAWs represent a promising and increasingly adopted approach for noncontact manipulation of biomaterials, facilitating precise cell patterning and assembly in bioprinting applications \cite{Wu2024}. Soft microgels offer a mechanically tunable platform to stabilize biofluid jets without relying on potentially cytotoxic additives or high-viscosity formulations. By enabling long, stable jets, our approach provides a path toward higher precision, reduced clogging, and improved structural fidelity in next-generation biofabrication systems.

Moreover, our results suggest broader applicability in soft matter and interfacial engineering. For example, in foams and emulsions, microgel-based interfacial networks could provide enhanced stability under fluctuating mechanical stresses, potentially delaying film rupture or coalescence. The interfacial scaffolding effect we describe may serve as a unifying design principle for creating far-from-equilibrium systems that resist deformation across multiple length and time scales.

Looking forward, several open questions remain. How do these soft interfacial networks ultimately fail under extreme strain? Can microgel-mediated stabilization be generalized to more complex fluids containing biomolecules or living cells? Could microgels be designed to mimic and reinforce fragile biological structures such as lipid bilayers \cite{Loi:2002PRE}, offering mechanical protection during physiological or processing stresses?

Ultimately, this work introduces a new paradigm for designing responsive and resilient fluid interfaces. Moving beyond conventional surfactant-based stabilization, we show that mechanically tunable particles can be used to control interfacial behavior under dynamic conditions. These insights not only deepen our understanding of capillary-driven instabilities but also offer practical strategies for advancing technologies in microfluidics, inkjet and bioprinting, and needle-free drug delivery. We anticipate that future efforts optimizing continuous liquid feeding, microgel softness, and driving frequencies will enable the generation of jets several times longer than currently achievable, expanding the horizons of nozzle-free jetting applications.


\section{Methods}

\subsection{SAW microfluidic chip}
This study employs standing surface acoustic waves (SAWs) to transform aqueous sessile droplets with varied volumes into a jet. Two counter-faced focused interdigital transducers (FIDTs) were used to generate counter-propagating SAWs with the same amplitude on the substrate where the sessile droplet is located. Interdigital transducers are structured on a 4" single-side polished lithium niobate wafer substrate with X-propagation direction (128$^\circ$YX LiNbO3), subsequently diced into single SAW jetting chips (8$\,\times\,$14$\,$mm$^2$). In the current chip layout, two focused interdigital transducers (IDT) ($60\,\mu$m wavelength, $30\,$degree focusing angle and matched to $50\,\Omega$ impedance) are opposing each other with a distance of $4\,$mm for SAW excitation based on superposition of two counter-propagating traveling SAWs. The construction and configuration of the SAW chip were described earlier \cite{winkler2017compact, winkler2015saw}. A signal generator (BSG F20, BelektroniG, Freital, Germany) was employed to drive the IDT with an excitation frequency of 64.4 MHz. During the experiments, the applied electrical power was maintained at approximately 4 W, with 2 W delivered to each IDT. To increase the static contact angle between the sessile drop and the substrate surface, SAW wafers were coated with a monolayer of 1H,1H,2H,2H-perfluorodecyltriethoxysilane (PFDTES) utilizing molecular self-assembly. The PFDTES coating method was previously disclosed \cite{Sablowski2022}. 
A microscopic image of a microfluidic SAW jetting chip with its components is shown in Fig.~\ref{fig:combi}b.

\subsection{Electrical and acoustical characterization}
The electrical radio-frequency (RF) behavior of the standing SAW chips was studied in the form of their complex scattering (S) parameters in a frequency range close to the Rayleigh-SAW excitation frequency. At the ideal working frequencies (|S11| minimum) and within a narrow bandwidth, i.e., 64.4$\pm$1 MHz, the focussing IDTs manufactured are ideally matched to 50 $\Omega$ impedance with a reflection coefficient of power of \(|S_{xx}|^2<2\%\)  and very low electric losses, characterized by an almost ideal baseline at\(|S11| > 0.95\). The SAW chips were placed in the chip holder and connected to an Agilent Technologies 5070B network analyzer via SMA cables and custom $50\,\Omega$-matched PCBs with electric waveguides and gold-coated spring pins. The network analyzer cables were calibrated up to the point where their male SMA connector met the PCBs female SMA connector.
Since the acoustic wave field is the dominant boundary condition for the SAW-liquid interaction, the wave field of a focused IDT structured on a SAW chip was measured around the jetting zone and for various frequencies close to the Rayleigh-SAW excitation frequency using a UHF 120 laser Doppler vibrometer (Polytec GmbH, Germany), mechanically and thermally stabilized for long-term measurements. Based on the results of traveling SAW wavefield measurements, the distance between the IDTs was carefully chosen to ensure that the center point of the sessile droplets, i.e. the intended jetting zone, corresponds to a defined SAW focal region.
\subsection{Sessile droplets}
 The aqueous sessile droplets with volumes ranging from 1 to 10 $\mu l$, were positioned on the SAW chip for jetting. The aqueous droplets contain PNIPAM microgels at varying concentrations (0.01 wt\%, 0.1 wt\%, and 1 wt\%). The PNIPAM microgels were synthesized by surfactant-free precipitation polymerization \cite{Pelton1986} with different cross-linker contents: 1 mol\%, 5 mol\%, and 10 mol\%, labeled as MG1, MG5, and MG10, respectively. It is noteworthy that microgels with lower cross-linker content exhibit greater softness \cite{Backes2018}, allowing them to stretch more at the interface. To support our experiments, we used surfactant (C$_{14}$TAB) solutions at two concentrations: 0.01 mM (0.00035 wt\%) and 1 mM (0.035 wt\%), corresponding to about 0.01 CMC and 0.3 CMC, respectively.
\subsection{Microgel characterisation}
Surface tension measurements (Fig. S2) were performed using a drop-shape analyzer OCA 20 (DataPhysics Instruments, Filderstadt, Germany). The dynamic viscosity of the microgel dispersions as a function of shear rate was measured with a rheometer (MCR702, Anton Paar). The results are shown in Fig. S3. The microgel sizes were measured with Dynamic Light Scattering (DLS) from LS Instruments (Switzerland) to obtain the hydrodynamic radius at different temperatures (Fig. S4). We used the Langmuir-Blodgett method (please see \cite{Picard2017}) to transfer microgels onto a silicon wafer at a certain surface pressure. The microgel distribution on the interface was then characterized using Atomic Force Microscopy (Cypher AFM system, Asylum Research, Santa Barbara, CA, USA). The corresponding results are presented in Fig.~\ref{fig:langmuir}.
\subsection{Optical imaging setup}
Two high-speed , one at TU Darmstadt (UX50 mini, FASTCAM, Photron, Japan) and one at SAWLab-Saxony (Phantom VEO 410, Vision Research Inc.), together with illumination systems were used to monitor the jet formation and development. The filming speed is 4 kfps. 
\subsection{Computer simulations}
To investigate the molecular-scale behavior of microgels at the air–water interface, we developed a coarse-grained dissipative particle dynamics (DPD) framework~\cite{Espanol:2017perspective,Wang:2020modeling}. In this hybrid polymer–solvent model, each microgel is represented as a cross-linked polymer network composed of monomer and cross-linker beads. All beads interact via the Weeks–Chandler–Andersen (WCA) potential~\cite{Weeks:1971JCP} to enforce excluded volume,
\begin{equation}
V_{\mathrm{WCA}}(r) =
\begin{cases}
4\epsilon\left[\left(\frac{\sigma}{r}\right)^{12} - \left(\frac{\sigma}{r}\right)^6\right] + \epsilon, & r \le 2^{1/6}\sigma, \\
0, & \text{otherwise},
\end{cases}
\end{equation}
where $\epsilon = k_{\mathrm{B}}T$ sets the energy scale and $r$ is the bead separation.  
Covalent bonds within the network are modeled by the finitely extensible nonlinear elastic (FENE) potential~\cite{Kremer:1990JCP},
\begin{equation}
V_{\mathrm{FENE}}(r) = -\epsilon k_{\mathrm{F}} R_0^2 \ln\left[1 - \left(\frac{r}{R_0\sigma}\right)^2\right], \quad r < R_0\sigma,
\end{equation}
with $k_{\mathrm{F}} = 15$ and $R_0 = 1.5$, ensuring structural integrity while allowing finite extensibility.

Following Ref.~\cite{Camerin:2019microgels}, the microgel network is generated using patchy particles: monomers carry two patches and cross-linkers carry four, otherwise behaving identically. This yields coarse-grained analogues of PNIPAM microgels. We simulate microgels with cross-linker concentrations $c = 2\%,\, 5\%,\, 10\%$ to match experimental conditions, with $N \approx 14{,}000$ monomers per microgel.

The surrounding water and air phases are represented by mesoscopic DPD beads interacting via conservative, dissipative, and random forces:
\begin{equation}
\vec{F}_{ij} = \vec{F}^C_{ij} + \vec{F}^D_{ij} + \vec{F}^R_{ij},
\end{equation}
where
\begin{align}
\vec{F}^C_{ij} &= a_{ij} \left(1 - \frac{r_{ij}}{R_c}\right) \hat{r}_{ij}, \\
\vec{F}^D_{ij} &= -\lambda w^2(r_{ij}) (\vec{v}_{ij} \cdot \hat{r}_{ij}) \hat{r}_{ij}, \\
\vec{F}^R_{ij} &= \Xi\, w(r_{ij})\, \theta_{ij}\, \hat{r}_{ij}.
\end{align}
Here $a_{ij}$ is the conservative interaction amplitude, $\lambda$ is the friction coefficient, $\Xi$ is the noise amplitude satisfying $\Xi^2 = 4\lambda k_{\mathrm{B}}T$, $w(r) = 1 - r_{ij}/R_c$ is the weight function, $\theta_{ij}$ is a Gaussian random variable of zero mean and unit variance, $\hat{r}_{ij}$ is the unit separation vector, and $\vec{v}_{ij}$ is the relative velocity. 

In the hybrid polymer–solvent model, water beads (w) and air beads (a) interact with microgel monomers (m) via DPD forces. The polymer–solvent parameter $a_{ms}$ controls solvent quality and thus the microgel swelling behavior~\cite{DelMonte2021}. To capture interfacial properties, we use a Flory–Huggins–based parameterization:  
$a_{ww} = a_{aa} = 8.8\,k_{\mathrm{B}}T/\sigma$ for water–water and air–air interactions, $a_{mw} = 4.5\,k_{\mathrm{B}}T/\sigma$ for polymer–water interactions, $a_{ma} = 5.0\,k_{\mathrm{B}}T/\sigma$ for polymer–air interactions, $R_c = 1.9\,\sigma$, and a dimensionless DPD bead density $\rho_{\mathrm{DPD}} = \rho R_c^3 = 4.5$~\cite{Gerelli:2024softness}. The molecular volumes of water and air are approximately $30 \,\mathrm{\AA}^3$ and $120 \,\mathrm{\AA}^3$, respectively~\cite{Wang:2020modeling}. To ensure equal effective DPD bead volumes, each DPD water bead represents two water molecules ($\approx 60 \,\mathrm{\AA}^3$), while each air molecule is mapped onto two DPD beads. This mapping yields an average DPD bead volume of $60 \,\mathrm{\AA}^3$ and the interaction range $R_c = 6.46 \,\mathrm{\AA}$. In the absence of microgels, these parameters yield an air–water surface tension $\gamma \approx 70 \,\mathrm{mN/m}$, consistent with experiment. Upon placing a microgel at the interface, we observe the characteristic “fried-egg” conformation.

To assemble interfacial monolayers, two microgels are placed at the air–water interface in a rectangular box with periodic boundaries. The system is equilibrated in explicit solvent, allowing microgels to adopt their interfacial morphology~\cite{Bochenek:2022situ}.  Due to periodicity, two air–water interfaces are present; we analyze the one occupied by the microgels by computing the virial pressure within $-40\sigma < z < 40\sigma$. The interfacial tension is obtained from the pressure anisotropy:
\begin{equation}
\gamma = L_z \left(P_{zz} - \frac{1}{2}(P_{xx} + P_{yy})\right),
\end{equation}
where $P_{zz}$ is the normal pressure, $P_{xx}$ and $P_{yy}$ are the lateral pressures, and $L_z$ is the sub-volume height.

All simulations are performed with LAMMPS~\cite{Thompson:2022lammps}. Units are given in $\sigma$ (length), $m$ (mass), $k_{\mathrm{B}}T$ (energy), $\tau_0 = \sqrt{m\sigma^2/k_{\mathrm{B}}T} \approx 1.3\times 10^{-12}\,\mathrm{s}$ (time), and $k_{\mathrm{B}}T/\sigma^2$ (surface tension).

\begin{acknowledgments} 
Fruitful discussions with Leslie Yeo from RMIT University are gratefully acknowledged. Financial support from the German Research Foundation (DFG)— Walter Benjamin project (460540240) and DFG-ANR Grant AERONEMS project (53301) is acknowledged.  Prof. von Klitzing is a member and A. Razavi is an associate member of RTG 2516 (Grant No. 405552959).

\end{acknowledgments}

\section{Author contributions}
A.R. (Amin Rahimzadeh), S.M., A.W., B.L. and R.v.K. designed this study. A.R. (Atieh Razavi), M.R., and A.R. (Amin Rahimzadeh) conducted the experiments. S.M. developed and performed the simulations. The scaling analysis is developed by S.M. and A.R. (Amin Rahimzadeh). A.R. (Atieh Razavi), M.R., S.M., A.W., B.L., R.v.K. and A.R. (Amin Rahimzadeh) interpreted and discussed the results. S.M., A.R. (Atieh Razavi), M.R. and A.R. (Amin Rahimzadeh) wrote the manuscript with the help of all other authors.

\section*{Code availability}
The computer code used for simulations is available from the corresponding authors upon request.

\section*{Data availability}
The data used for this paper are available from the corresponding authors upon request.

\section*{Movies}
Movies supporting the findings of this study are available at \url{https://tudatalib.ulb.tu-darmstadt.de/items/d0f9e159-0277-4cc9-91a2-d4ddcbc39fc9}.

\vspace{1cm}

1) \textbf{Jetting with pure water (Water.avi)}: A $10$-$\mathrm{\mu l}$ water droplet forms a jet under SAW excitation and breaks up after reaching its maximum length (10.82 mm).

\vspace{0.25 cm}

2) \textbf{Jetting with microgels (Microgel.avi)}: A $10$-$\mathrm{\mu l}$ soft microgel (MG1) droplet forms a jet under SAW excitation and breaks up after reaching its maximum length (15.59 mm).

\vspace{0.25 cm}

3) \textbf{Computer simulation of soft microgels (Soft.mp4)}: During elongation, two soft microgels (MG2) remain interconnected, maintaining continuous coverage of the air–water interface.

\vspace{0.25 cm}

4) \textbf{Computer simulation of stiff microgels (Stiff.mp4)}: Under identical conditions, two stiff microgels (MG10) lose contact and disentangle, exposing bare air–water interface.

\providecommand{\noopsort}[1]{}\providecommand{\singleletter}[1]{#1}%
%


\ifarXiv
    \foreach \x in {1,...,\numbersupplementpages}
    {
        \clearpage
        \includepdf[pages={\x}]{\supplementfilename}
    }
\fi

\end{document}